\documentclass[aip,apl,reprint,showpacs]{revtex4-1}
\usepackage{verbatim}
\usepackage{textcomp}
\usepackage{graphicx}

\begin{document}


\title{Electron temperature in electrically isolated Si double quantum dots} 

\affiliation{Hitachi Cambridge Laboratory, J.J. Thomson Avenue, Cambridge, CB3 0HE, U.K.}
\author{A. Rossi}\email[Electronic mail: ]{ar446@cam.ac.uk}
\author{T. Ferrus}


\author{D.A. Williams}



\date{\today}

\begin{abstract}
Charge-based quantum computation can be attained through reliable control of single electrons in lead-less quantum systems. Single-charge transitions in electrically-isolated double quantum dots (DQD) realised in phosphorus-doped silicon can be detected via capacitively coupled single-electron tunnelling devices. By means of time-resolved measurements of the detector's conductance, we investigate the dots' occupancy statistics in temperature. We observe a significant reduction of the effective electron temperature in the DQD as compared to the temperature in the detector's leads. This sets promises to make isolated DQDs suitable platforms for long-coherence quantum computation.
\end{abstract}


\maketitle 
\begin{figure}[]
\includegraphics[scale=0.43]{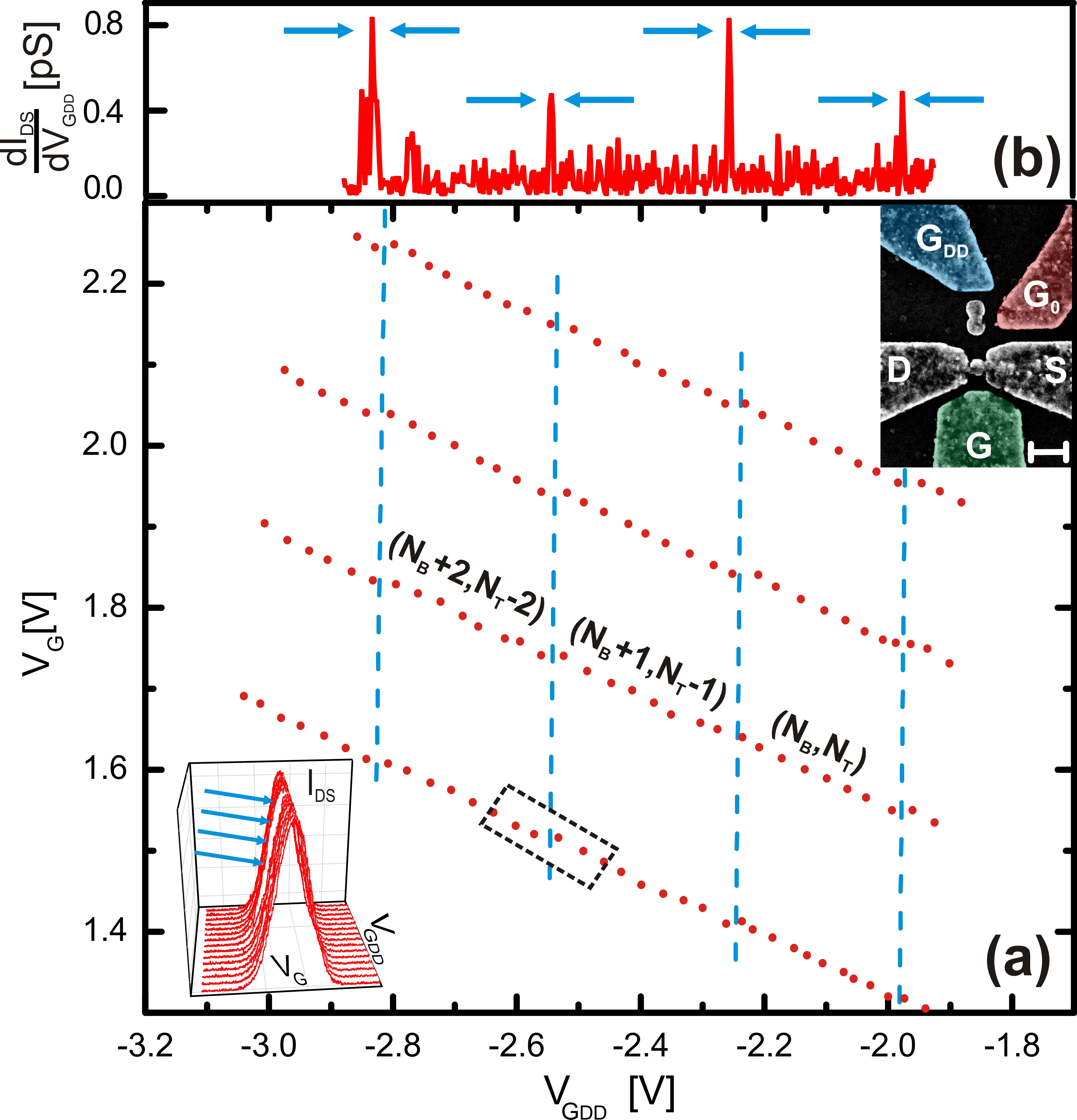}
 \caption{\label{fig:maxima}(a) Positions of the detector's Coulomb peak maxima as a function of $V_G$ and $V_{G_{DD}}$ (dotted line). Charge transitions in the DQD are sensed as shifts in the position of the peak maxima that periodically break their trajectories. Dashed lines highlight the loci of the shifts as parallel straight lines that separate regions of equal charge distribution in the DQD. Top inset: false colour SEM micrograph of a representative device with scale bar of 200 nm. Designation of electrodes: detector's source (S), drain (D) and gate (G); DQD's gates (G$_\mathsf{DD}$, G$_0$). Bottom inset: SET current in the region highlighted by the dashed rectangle in the main plot. Current range: 0pA$\leq I_{DS}\leq$55pA. Arrows highlight the peaks' shift. (b) Differential conductance of the detector as a function of $V_{G_{DD}}$. $V_G$ is simultaneously compensated to keep the conductance level at a fixed value. Pairs of arrows highlight sharp spikes which reveal sudden change of conductance due to tunnelling events in the DQD.}
 \end{figure}
\begin{figure}[b]
\includegraphics[scale=0.62]{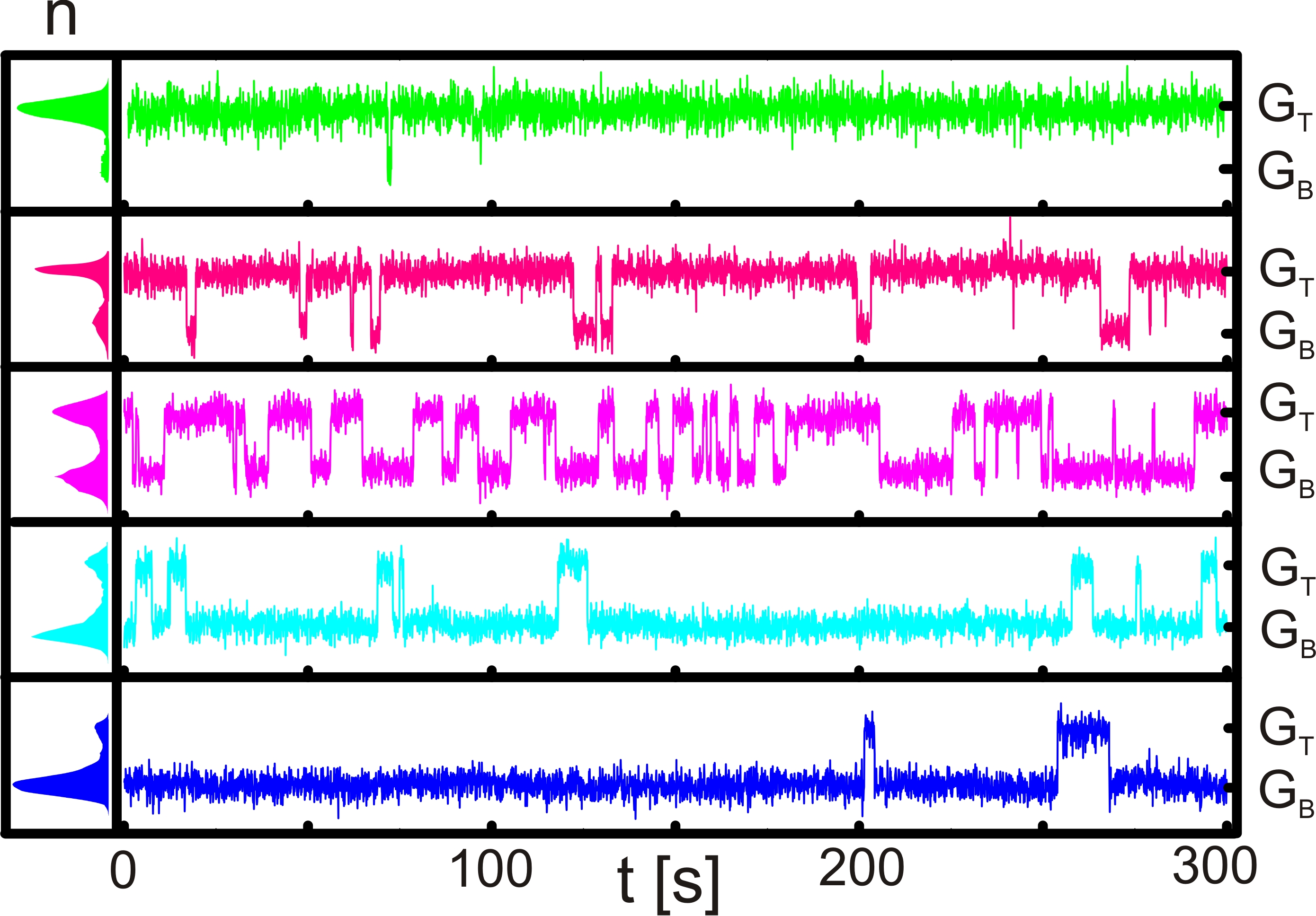}
\caption{\label{fig:time} Time-resolved traces of the normalised SET conductance near a transition point for $V_{G_{DD}}$=-2.549V, -2.551V, -2.553V, -2.555V, -2.558V from top to bottom. The conductance fluctuates between two discrete levels according to the DQD occupancy modification given by the gate voltage applied. Histograms on the left hand side account for the time the system spends in each state; $n$ is the normalised number of points per trace at the conductance level given on the vertical axis.}
\end{figure}
\begin{figure}[]
\includegraphics[scale=0.45]{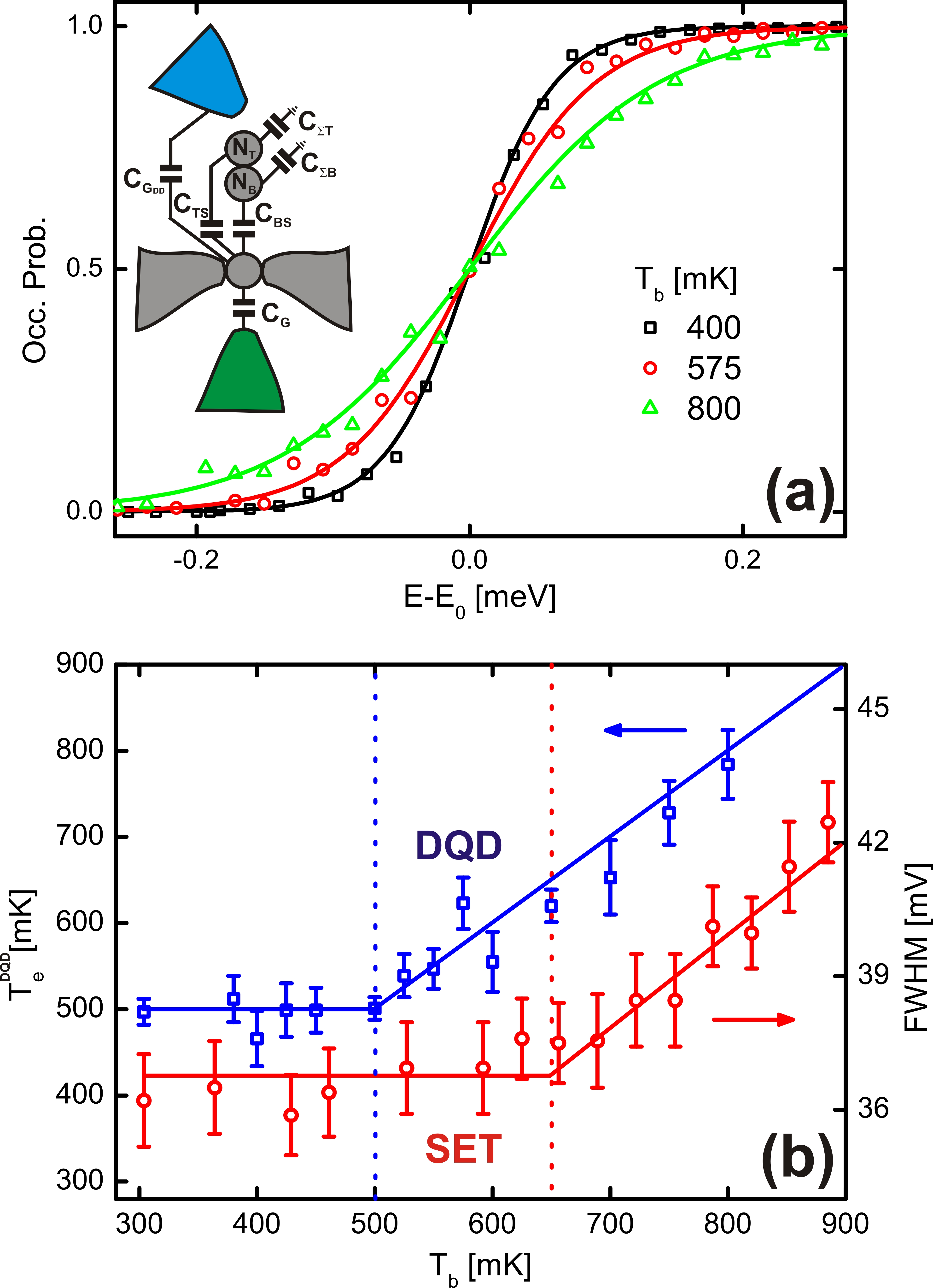}
 \caption{\label{fig:distr} (a) Occupation probability function extracted from time-resolved traces as a function of the electrochemical potential relative to $E_0$. Data points are taken at different base temperatures as reported in the legend. Solid lines are best fits according to the FD distribution. Left inset: equivalent circuit indicating the main coupling capacitances to the detector as well as the self-capacitances of the two dots. Coupling to gate G$_0$ is omitted. (b) Estimates of electron temperatures in the DQD (squares) and SET (dots). Data points for the DQD are obtained by using  $T^{DQD}_e$ as a fitting parameter for distributions of the type shown in (a). FWHM of the SET is evaluated by fitting Coulomb resonances with Lorentzian peak functions. Dashed lines indicate the knee points which estimate the minimum electron temperatures in the detector, $T^{det}_{e}\approx$~650~mK, and the DQD, $T^{DQD}_{e}\approx$~500~mK.}
 \end{figure}

Quantum mechanical charge and spin states of electrons confined in semiconductor double quantum dots (DQD)~\cite{hanson,haya,peter}~have recently attracted much interest, as they can be exploited to implement solid-state quantum computation. One key requirement to perform quantum logic operations is a long coherence time for the qubit-embodying states. Among other system materials, silicon is particularly suited to retain spin-coherence for long time mainly due to the existence of a stable isotope ($^{28}$Si) without nuclear magnetic moment.~\cite{tyry} Another approach to mitigate the decoherence introduced by the interaction with the environment is the suppression of exchange processes with electrons in the reservoirs.~\cite{yurke} This has proven to be beneficial if a charge-qubit implementation is to be preferred. Indeed, single-qubit operations have been successfully implemented in trench-isolated silicon double quantum dots,~\cite{gorman} where electrons are confined in a lead-less system and only capacitively coupled to gates.\\\indent
Here, we investigate the electronic occupation of phosphorus-doped isolated DQDs using a single-electron tunnelling device (SET) as a charge sensor. By means of time-resolved charge detection, the probability of occupation of each dot is evaluated as the control gate is swept across a degeneracy point (i.e. alignment of energy levels which leads to electron tunnelling). From these measurements we have extracted the electron temperature in the isolated quantum system and compared it with the one measured in the SET leads. The advantageous effects of a lead-decoupled system are quantitatively observed as a significant temperature reduction in the DQD.\\\indent
The top inset of Fig.~\ref{fig:maxima}(a) shows a scanning electron micrograph (SEM) image of a representative device. The use of silicon-on-insulator wafers ensures spatial confinement of electrons in the vertical direction; reactive ion etching is used to define deep trenches which produce confinement within the horizontal plane. The silicon active layer is doped with a concentration of phosphorus atoms of about 3$\times$10$^{19}$ cm$^{-3}$ to provide the nano-structure with free carriers. From the density of implanted donors, we estimate that a 40 nm diameter dot contains about 2000 electrons. Both electron-beam and optical lithography are employed to define the device pattern. Full details of the fabrication process, tunnel barriers formation and influence of interface trapping for this system are reported elsewhere.~\cite{myJAP} In order to keep the operating point of the detector in the most sensitive region while inducing transitions in the DQD, gates G and G$_\mathsf{DD}$ are simultaneously swept with a fixed ratio.  By contrast, gate G$_0$ is always kept at a fixed voltage. Experiments have been performed in a $^3$He cryostat at the base temperature $T_b=$~295 mK, unless otherwise stated.\\\indent
Fig.~\ref{fig:maxima}(a) shows the positions of the detector's Coulomb peak maxima as a function of the voltage applied to gates G$_\mathsf{DD}$ and G. The plot reveals the occurrence of Coulomb Blockade (CB) in the SET; the resonances' slope indicates that both gates are coupled to the SET island and significantly affect its potential. In particular, it is possible to determine the relative strength with which each gate is coupled to the SET in terms of gate capacitance ratio. Indeed, the slope of the CB features ($s=\frac{|\Delta V_{G_{DD}}|}{|\Delta V_G|}=2.8$) directly reflects the ratio between the SET gate capacitances (see left inset of Fig.~\ref{fig:distr}), which can be written as $C_G=s\cdotp C_{G_{DD}}$; this defines the voltage compensation ratio needed to keep the SET at a fixed bias point. Most interestingly, we observe that the trajectory of the conductance peaks in [$V_G$,$V_{G_{DD}}$] space is nearly periodically shifted. In the bottom inset, the shift highlighted by the dashed rectangle is shown as a misalignment between two adjacent groups of peaks in two-gate voltage space. As previously reported,\cite{myAPL} this effect is the clear signature of single-charge tunnelling within the nearby DQD structure; in particular, for more negative $V_{G_{DD}}$ we expect electrons to move from the top dot (T) to the bottom one (B). Each of these events produces a shift of about 1$\%$ of a CB period which is equivalent to an induced charge in the SET island of 0.01$e$, where $e$ is the elementary charge. This directly depends on the amount of mismatch in the coupling of each dot to the detector, as it has been found~\cite{myAPL} that $\Delta\propto C_{TS}/C_{\Sigma_T}-C_{BS}/C_{\Sigma_B}$, where $\Delta$ is the modification of the SET island potential upon a tunneling event in the DQD, $C_{TS}$($C_{BS}$) is the coupling capacitance between the top (bottom) dot and the SET island and $C_{\Sigma_T}$ ($C_{\Sigma_B}$) is the total self-capacitance of the top (bottom) dot. It is worth noting that the shifts have all nearly equal amplitude (1$\%$) and draw parallel lines (see dashed lines in the main plot) that divide the voltage space in regions of stable charge distribution within the DQD. This symmetry indicates that the transitions originate all from the same spatial location (the DQD) which has a fixed capacitive coupling to both the charge sensor and the gate electrodes. Fig.~\ref{fig:maxima}(b) depicts the differential conductance of the detector as the gate voltages $V_{G_{DD}}$ and $V_G$ are mutually compensated along the trajectory followed by the lower CB resonance in (a) to keep maximum charge sensitivity. Sharp resonances clearly appear in correspondence of the voltage shifts; this reflects, once more, the occurrence of discrete changes in the electrostatic potential of the SET due to remote charge transfer.\\\indent 
In order to observe the time-dependent evolution of the DQD near a transition, traces are recorded for different gate voltages across a degeneracy point. In Fig.~\ref{fig:time} we report five representative traces of the SET conductance for time intervals of 300~s at a sampling frequency of 13 Hz. Data are taken in the vicinity of the transition shown in the inset of Fig.~\ref{fig:maxima}(a). It is evident that near the degeneracy point the system switches randomly between the two states associated with two discrete conductance levels of the SET (i.e. $G_T$ and $G_B$ for an electron occupying the top and the bottom dot, respectively). The fraction of time that the system spends in each state is a function of the gate voltage and approaches 50$\%$ at the degeneracy point (central trace). Such an information can be extracted from the time-dependent plots by counting the number of points, $n$, per conductance bin. A histogram of $n$ versus conductance is reported next to each relevant trace. The histograms allow one to directly visualize how gate voltages control the crossover from localised to delocalised charge states.\\\indent
The presence of a range of gate voltages for which the system is degenerate (as opposed to sharp transitions at a single-valued voltage) suggests that tunnelling is affected by inelastic scattering processes mainly due to electron-phonon interactions. The gate voltage spread across the transition can be used to extract information on the energy level broadening and the effective electron temperature in the isolated DQD structure. From histograms similar to those shown in Fig.~\ref{fig:time}, the distribution functions of Fig.~\ref{fig:distr}(a) can be built by converting the gate voltage into energy scale,
\begin{equation}
\label{eq:alpha}
E-E_0=-\alpha_{DQD}(V_{G_{DD}}-V_{G_{DD_0}})
\end{equation}
where $\alpha_{DQD}$ is the lever arm of gate G$_\mathsf{DD}$ to the DQD, $E_0$ and $V_{G_{DD_0}}$ are the degeneracy point electrochemical potential and voltage, respectively. Following the theory by Beenakker,~\cite{beenak} we can treat the discrete energy spectrum in the DQD as a continuum, given that for our system $E_C\gg k_BT^{DQD}_e\gtrsim\Delta\epsilon$, $k_B$ being the Boltzmann's constant, $T^{DQD}_{e}$ the electron temperature in the DQD, $\Delta\epsilon$ the single-particle level separation and $E_C$ the dots' charging energy. Under this assumption, the distribution function can be approximated by the Fermi-Dirac (FD) distribution,
\begin{equation}
f(E-E_0)=\frac{1}{1+e^{-(E-E_0)/k_BT^{DQD}_{e}}}
\end{equation}
from which the effective electron temperature can be extracted. It is worth pointing out that our system largely satisfies the classical limit approximation as $E_C\approx$~17 meV, $k_BT^{DQD}_e\approx$~30 $\mu$eV and $\Delta\epsilon\approx$~20 $\mu$eV. Fig.~\ref{fig:distr}(a) depicts three representative distributions for different values of $T_b$. Solid line fits allow one to estimate the values of $T^{DQD}_{e}$ which appear not to be affected by the base temperature up to approximately 500 mK, as shown in Fig.~\ref{fig:distr}(b) (squared data points). Indeed, it is clear from the fitting lines in blue that for $T_b\lesssim$~500~mK, $T^{DQD}_{e}\approx$~500~mK whereas for  $T_b>$~500~mK, $T_b\approx$~$T^{DQD}_{e}$. This is the evidence that the lowest limit for the electron temperature in the DQD is set to roughly 500 mK by our cryogenic set-up. In order to correctly extract the electron temperature from the occupancy distributions, it is paramount to know precisely the value of  $\alpha_{DQD}$ to be used in Eq.~(\ref{eq:alpha}). Given that direct transport measurements are not allowed due to the isolated nature of the DQD, the estimate of  $\alpha_{DQD}$ has been attained as follows. Initially, we have used a value of first attempt by assuming  $\alpha_{DQD}=\alpha_{SET}=$~0.033~eV/V, where $\alpha_{SET}$ is the lever arm of gate G to the SET. This can be easily extracted from the diamond plot of the detector. Next, from the observation that  $T^{DQD}_{e}$ is nearly constant up to  $T_{b}\approx$~500 mK, we have fitted the occupancy distribution at $T_{b}=$~500 mK using  $\alpha_{DQD}$ as a fitting parameter and  $T^{DQD}_{e}=$~500 mK as a fixed parameter. The final value of the lever arm so obtained is  $\alpha_{DQD}=$~0.055\textpm 0.001~eV/V. The fact that the line of best fit in Fig.~\ref{fig:distr}(b) for $T_b>$~500~mK satisfies the condition $T_b=$~$T^{DQD}_{e}$ is a decisive consistency check which corroborates the approach used to evaluate the lever arm.\\\indent
It is now interesting to compare the value of minimum electron temperature for the DQD and the detector's leads.  To this end, we measure the full width at half maximum (FWHM) of the Coulomb peaks in the SET as the base temperature of the cryostat is increased. We find that the width scales linearly with temperature~\cite{foxman} from a minimum value $T^{det}_{e}\approx$~650 mK, which is an estimate of the electron temperature in the detector (see dotted data points in Fig.~\ref{fig:distr}(b)). The discrepancy between the two electron temperatures is a convicing manifestation of the beneficial effects of decoupling quantum systems from leads electrically connected to room temperature electronics. However, the presence of other concurrent heating mechanisms in the DQD is indicated by the fact that the lowest $T^{DQD}_{e}$ is larger than any base temperature value below 500 mK, which is unexpected for an isolated system in thermal equilibrium with the phonon bath. A number of reasons may be the cause of this effect. Firstly, the presence of the charge detector in the vicinity of the DQD may produce heating exchange due to phonon-mediated back-action.~\cite{gas} In other words, the current flowing through the SET produces ohmic dissipation in the leads that can be absorbed by the phonon bath and transferred to the electrons in the DQD. If this mechanism turned out to be dominant, we would observe modifications in the value of $T^{DQD}_e$ as a function of the detector's bias point. This will be the focus of further investigations. Moreover, the weak coupling between the electrons and the phonon bath, which is typically in place in nano-scale systems because of finite-size effects,~\cite{tilke,ziman}~should be considered. This may lead to exceedingly long thermalisation times on the scale of the time-resolved experiments presented here and consequently result in an increase of the electron temperature. Finally, a temperature difference may also arise due to some difficulty in attaining premium thermal conduction between the condensed $^3$He and the cold finger where the device is plugged in.\\\indent
In conclusion, this work has shown that control of charge state can be achieved in fully electrically isolated DQD systems. Most interestingly, we have demonstrated that the benefit of decoupling the DQD from the leads can be quantitatively estimated as a strong reduction in electron temperature. Given that the mitigation of decoherence sources is a crucial requisite for reliable quantum manipulations, this system promises to be a viable candidate for charge-based quantum computation.\\\indent
This work was partly supported by Project for Developing Innovation Systems of the Ministry of Education, Culture, Sports, Science and Technology (MEXT), Japan.


%
%





\begin{thebibliography}{15}
\bibitem{hanson}
R.~Hanson, L.~P. Kouwenhoven, J.~R. Petta, S.~Tarucha, and L.~M.~K.
  Vandersypen.
\newblock {\em Rev. Mod. Phys.}, 79:1217--1265, 2007.

\bibitem{haya}
T.~Hayashi, T.~Fujisawa, H.~D. Cheong, Y.~H. Jeong, and Y.~Hirayama.
\newblock {\em Phys. Rev. Lett.}, 91:226804, 2003.

\bibitem{peter}
K.~D. Petersson, J.~R. Petta, H.~Lu, and A.~C. Gossard.
\newblock {\em Phys. Rev. Lett.}, 105:246804, 2010.

\bibitem{tyry}
A.M. Tyryshkin, J.J.L. Morton, S.C. Benjamin, A.~Ardavan, G.A.D. Briggs, J.W.
  Ager, and S.A. Lyon.
\newblock {\em J. Phys.: Condens. Matter}, 18:S783, 2006.

\bibitem{yurke}
I.V. Yurkevich, J.~Baldwin, I.V. Lerner, and B.L. Altshuler.
\newblock {\em Phys. Rev. B}, 81:121305, 2010.

\bibitem{gorman}
J.~Gorman, D.~G. Hasko, and D.~A. Williams.
\newblock {\em Phys. Rev. Lett.}, 95:090502, 2005.

\bibitem{myJAP}
A.~Rossi and D.G. Hasko.
\newblock {\em J. Appl. Phys.}, 108:034509, 2010.

\bibitem{myAPL}
A.~Rossi, T.~Ferrus, G.~J. Podd, and D.~A. Williams.
\newblock {\em Appl. Phys. Lett.}, 97(22):223506, 2010.

\bibitem{beenak}
C.~W.~J. Beenakker.
\newblock {\em Phys. Rev. B}, 44:1646--1656, 1991.

\bibitem{foxman}
E.~B. Foxman, U.~Meirav, P.~L. McEuen, M.~A. Kastner, O.~Klein, P.~A. Belk,
  D.~M. Abusch, and S.~J. Wind.
\newblock {\em Phys. Rev. B}, 50:14193--14199, 1994.

\bibitem{gas}
U.~Gasser, S.~Gustavsson, B.~K\"ung, K.~Ensslin, T.~Ihn, D.~C. Driscoll, and
  A.~C. Gossard.
\newblock {\em Phys. Rev. B}, 79:035303, 2009.

\bibitem{tilke}
A.~Tilke, L.~Pescini, A.~Erbe, H.~Lorenz, and R.~Blick.
\newblock {\em Nanotech.}, 13:491, 2002.

\bibitem{ziman}
J.~Ziman.
\newblock {\em Electrons and phonons}.
\newblock Clarendon, Oxford, 1960.
\newblock p.307.

\end{thebibliography}

\end{document}